\documentclass[english, 11pt, a4paper]{article}

\usepackage{slashed}
\usepackage[utf8]{inputenc}
\usepackage{babel}
\usepackage{graphics}
\usepackage{graphicx}
\usepackage{texdraw}
\usepackage{epsfig}
\usepackage{float}
\usepackage{amsmath}
\usepackage{amssymb}
\usepackage{enumerate}
\usepackage[all]{xy}
\usepackage{fullpage}
\usepackage[small]{caption}
\usepackage{hyperref}

\newcommand{\be}{\begin{equation}}
\newcommand{\ee}{\end{equation}}

\begin{document}

\title{Topology in the $SU(N_f)$ chiral symmetry restored phase of unquenched QCD and
  axion cosmology II}

\date{}
\author{Vicente ~Azcoiti \\
        Departamento de F{\'i}sica Te\'orica, Facultad de Ciencias \\
        Universidad de Zaragoza, Pedro Cerbuna 9, 50009 Zaragoza, Spain\\}

\maketitle

\begin{abstract}

  We investigate the physical consequences of the survival of the effects of the $U(1)_A$
  anomaly in the chiral symmetric phase of $QCD$, and show that the free
  energy density is a singular function of the quark mass $m$, in the chiral limit,
  and that the $\sigma$ and $\bar\pi$ susceptibilities diverge in this limit at any
  $T\ge T_c$. We also show that the difference between the
  $\bar\pi$ and $\bar\delta$ susceptibilities diverges in the chiral limit at any
  $T\ge T_c$, a result which seems to be excluded by recent results of Tomiya et al.
  from numerical simulations of two-flavor QCD. We also discuss on the generalization of
  these results to the $N_f\ge 3$ model.

\end{abstract}

\vfill\eject

\section{Introduction}

Quantum field theories with a topological term in the action \cite{vicari} have proved to
be particularly
challenging to investigate. The strong $CP$ problem in strong interactions, and the Haldane
conjecture and the quantum Hall effect in condensed matter physics are representative 
of important open problems in theoretical physics closely related to the topological properties
of the model.

In what concerns $QCD$, understanding the role of the $\theta$ parameter and its connection with
the strong CP problem is a major challenge. On the other hand the aim to elucidate the existence
of new low-mass, weakly interacting particles from a theoretical, phenomenological and
experimental point of view, is intimately related to this issue. The light particle that has
mostly gathered attention is the axion, predicted by
Weinberg and Wilczek \cite{weinberg}, \cite{wilczek} in the Peccei and Quinn mechanism
\cite{pq} to explain the absence 
of parity and temporal invariance violations induced by the QCD vacuum. The axion is also one of
the more interesting candidates to make the dark matter of the universe, and the axion potential
plays a fundamental role in the determination of the dynamics of the axion field. 
Moreover, the way in which the $U(1)_A$ anomaly manifests itself in the chiral 
symmetry restored phase of $QCD$ at high temperature could be tested when probing the 
$QCD$ phase transition in relativistic heavy ion collisions. 

The topological properties of the $QCD$-vacuum are intrinsically nonperturbative,
thus requiring a nonperturbative
approach. The calculation of the topological susceptibility by means of simulations in lattice
$QCD$ is already a challenge, but calculating the complete potential requires a strategy to deal
with the so-called sign problem, that is, the presence of a highly oscillating term in
the path integral, which prevents the applicability of the importance sampling
method \cite{vicari}. But the $QCD$ axion model relates the topological susceptibility $\chi_T$
at $\theta=0$ with
the axion mass $m_a$ and decay constant $f_a$ through the relation $\chi_T = m^2_a f^2_a$, and 
the axion mass is an essential ingredient in the calculation of the
axion abundance in the Universe. Therefore a precise computation of the temperature
dependence of the topological susceptibility in $QCD$ becomes of primordial interest in
this context. Indeed, several calculations of this quantity 
in unquenched $QCD$ have been recently reported \cite{martinelli},
\cite{petre}, \cite{javier}.

Unfortunately there are strong discrepancies among these three calculations. Bonati et al. 
\cite{martinelli} explore $N_f=2+1$
$QCD$ in a range of temperature going from $T_c$ to around $4T_c$, and their results for
the topological susceptibility differ strongly, both in size and in temperature
dependence, from the dilute instanton gas prediction, giving rise to a shift of the axion
dark matter window of almost one order of magnitude with respect to the instanton
computation. Petreczky et al. \cite{petre} observe however 
very distinct temperature dependences of the topological susceptibility in the ranges above
and below 250 MeV: while for temperatures above 250 MeV, the dependence is found to be
consistent with the dilute instanton gas approximation, at lower temperatures the fall-off of
topological susceptibility is milder.  Borsanyi et al. \cite{javier} find, on the other hand, a 
topological susceptibility many orders of magnitude smaller than that of
reference \cite{martinelli} in the cosmologically relevant temperature region. These discrepancies
among the three calculations make more interesting, if possible, a theoretical 
approach to the issue.

The absence of the typical effects of the
$U(1)_A$ anomaly in the chiral symmetry restored phase of $QCD$ at high-temperature was
suggested in \cite{cohen1}, \cite{cohen2}, and investigated later on 
\cite{cuatro}-\cite{ding}. Indeed, years ago Cohen \cite{cohen1} showed, using the
continuum formulation of two flavor $QCD$, and 
assuming the absence of the zero mode's
contribution, that all the disconnected contributions to
the two-point correlation functions in the $SU(2)_A$ symmetric phase
at high-temperature vanish in the chiral limit. The main conclusion of this work was that
the eight scalar and pseudoscalar mesons should have the same mass
in the chiral limit, the typical effects of the $U(1)_A$ anomaly being absent in this phase.
Furthermore he argued in \cite{cohen2} that the analyticity of the free energy density in the
quark mass $m$ around $m=0$, in the high temperature phase, imposes constraints on the
spectral density of the Dirac operator around the origin which are enough to guarantee the
previous results. Later on Aoki et al. \cite{diez} got constraints on the Dirac spectrum of
overlap fermions, strong enough for all of the
$U(1)_A$ breaking effects among correlation functions of scalar and pseudoscalar
operators to vanish, and they concluded that there is no remnant of the $U(1)_A$ anomaly above
the critical temperature of two flavor $QCD$, at least in these correlation functions. Their
results were obtained under the assumptions that $m$-independent observables are analytic
functions of the quarks mass $m$, at $m=0$, and that the Dirac spectral density can be expanded
in Taylor series near the origin, with a nonvanishing radius of convergence. 

More recently we investigated by analytical methods in reference \cite{trece}  
the topological
properties of $QCD$ in the high temperature chiral symmetric phase, and we
summarize here what was the starting hypothesis in \cite{trece}, 
its physical motivation,
and the main conclusion which follows from it. The starting hypothesis was to assume
that the perturbative expansion of the free energy density in powers of the quark
mass, $m$, has a nonvanishing convergence radius in the high temperature chiral
symmetric phase of $QCD$. This is just what we expect on physical grounds if all 
correlation lengths remain
finite in the chiral limit, and the spectrum of the model shows therefore a mass gap
also in this limit. The main conclusion which followed from this hypothesis was
that all the topological effects of the axial anomaly should disappear in this phase,
the topological susceptibility and all $\theta$-derivatives of the free energy
density vanish, and the theory becomes $\theta$ independent at any
$T > T_c$ in the infinite-volume limit.

Accordingly, the free energy density should be a
singular function of the quark mass, in the chiral limit, if the topological effects
of the $U(1)_A$ anomaly survive in the chiral symmetry restored phase of $QCD$ at finite
temperature, and the main purpose of this article is to investigate this issue.
Our starting hypothesis will be now that the topological effects of the anomaly
survive in the high temperature phase of $QCD$, and the model shows therefore a
nontrivial $\theta$-dependence in this phase. Under this assumption we will show
here that indeed the free energy density is a singular function of the quark mass, $m$,
in the chiral limit at any $T>T_c$, and that the correlation length and the
$\sigma$ and $\bar\pi$
susceptibilities diverge in this limit, as well as the difference between the
$\bar\pi$ and $\bar\delta$ susceptibilities.

We will show first this result following the line of argumentation developed 
in \cite{trece}, and thereafter exploiting the qualitative features of the phase 
diagram of $QCD$ in the $Q=0$ topological sector. Our main conclusion is that this scenario
should be excluded, in the $N_f\ge 3$ case, by universality and renormalization group 
arguments, and in the two flavor model, by recent results of numerical simulations of high
temperature two-flavor $QCD$ \cite{catorce}\footnote{We should notice however that the results
  of reference \cite{catorce} for the two-flavor case do not agree with those of reference
  \cite{sharma}, and any further clarification on this point would be therefore welcome.}.

\section{$\sigma$ and $\eta$ susceptibilities}

The Euclidean continuum Lagrangian of $N_f$ flavors $QCD$ with a $\theta$-term is

\begin{equation}
L = \sum_f L^f_F + \frac{1}{4} F^a_{\mu\nu}\left(x\right)F^a_{\mu\nu}\left(x\right)
+ i\theta \frac{g^2}{64\pi^2} \epsilon_{\mu\nu\rho\sigma}
F^a_{\mu\nu}\left(x\right)F^a_{\rho\sigma}\left(x\right)
\label{eulagran}
\end{equation}
with $L^f_F$ the fermion Lagrangian for the $f$-flavor, and

\begin{equation}
Q = \frac{g^2}{64\pi^2} \int d^4x\hskip 0.1cm \epsilon_{\mu\nu\rho\sigma}
F^a_{\mu\nu}\left(x\right)F^a_{\rho\sigma}\left(x\right)
\label{ftopcharg}
\end{equation}
is the topological charge of the gauge configuration.

To avoid ultraviolet divergences we will assume along this paper a lattice 
regularization, the Ginsparg-Wilson (G-W) fermions \cite{Ginsparg}, 
from which the overlap fermions \cite{Neuberger} are an explicit realization,
which shares with the continuum all essential ingredients and gives at the
same time mathematical rigor to all developments. Indeed G-W fermions have a
$U(1)_A$ anomalous symmetry \cite{Luscher}, good chiral properties, a
quantized topological charge, and allow us to establish and exact index
theorem on the lattice \cite{Victor}. 

The partition function of the model can be written as a sum over all topological 
sectors, $Q$, of the partition function in each topological sector times a 
$\theta$-phase factor, as follows

\begin{equation}
Z(\theta) = \sum_{Q} Z_Q e^{i\theta Q}
\label{zeta}
\end{equation}
where Q, which takes integer values, is bounded at finite volume by the number of
degrees of freedom. 
At large spatial lattice volume $V_x$ the partition function should behave as

\begin{equation}
Z\left(\theta\right) = e^{-V_x L_t E\left(\beta,m,\theta\right)}
\label{zetalarge}
\end{equation}
where $E\left(\beta,m,\theta\right)$ is the free energy density, $\beta$ the inverse 
gauge coupling, $m$ the quark mass, 
and $L_t$ the lattice 
temporal extent or inverse physical temperature $T$. Moreover the mean 
value of any intensive operator $O$, as for instance the scalar and pseudoscalar
condensates, or any correlation function, in the $Q=0$ topological sector, can be
computed as 

\begin{equation}
\left< O\right>_{Q=0} = \frac{\int d\theta \left< O\right>_\theta Z(\theta, m)}
{\int d\theta Z(\theta,m)}
\label{mascurioso}
\end{equation}
with $\left< O\right>_\theta$ the mean value of $O$ computed with the integration 
measure (\ref{eulagran}).
 
We are also assuming along this paper that the topological effects of the $U(1)_A$ anomaly
survive in the high temperature phase of $QCD$, or in other words, that the free energy 
density (\ref{zetalarge}) shows a non trivial $\theta$-dependence also in the high 
temperature chiral symmetric phase. Then, since the free energy density, as a 
function of $\theta$, has its absolute minimum at $\theta=0$ for non-vanishing quark 
masses, the following relation holds in the infinite lattice volume limit

\begin{equation}
\left< O\right>_{Q=0} = \left< O\right>_{\theta=0}
\label{mascuriosob}
\end{equation}
We want to remark that, as discussed in \cite{trece}, in spite of the fact that the 
$Q=0$ topological sector is free from the $U(1)_A$ global anomaly, and spontaneously breaks 
the $U(N_f)_A$ axial symmetry at $T=0$, equation (\ref{mascuriosob}) is 
compatible with a massive flavor-singlet pseudoscalar meson in the chiral limit. 
We will also make use of equation (\ref{mascuriosob}) along this paper.

Let us consider, for simplicity, the two-flavor model with degenerate up and down 
quark masses. In the high temperature phase 
the $SU(2)_A$ symmetry is fulfilled in the ground state for massless quarks, and therefore 
the mean value of the flavor singlet scalar condensate $\left< S\right>$, as well as 
of any order parameter 
for this symmetry, vanishes in the chiral limit. 
Moreover the infinite lattice volume limit and the chiral limit should commute, 
provided the order parameter remains bounded. In addition equation 
(\ref{mascuriosob}) implies that the $SU(2)_A$ symmetry is also fulfilled in the 
$Q=0$ topological sector. However, the $U(1)_A$ symmetry should be spontaneously 
broken in this sector, giving account in this way for the $U(1)_A$ anomaly\footnote{The 
  Goldstone theorem however can be fulfilled without a Nambu-Goldstone boson \cite{trece}.}.
In fact, the $\sigma$ 
and $\eta$ correlation functions, which take in the $Q=0$ sector the same value as in
$QCD$ at $\theta=0$ (\ref{mascuriosob}), 
should be different in the chiral 
limit, and the difference of these correlation functions is an order parameter for the
$U(1)_A$ symmetry of the $Q=0$ sector. Therefore 
we can characterize the ground states of the $Q=0$ sector, in the chiral limit, 
by an angle $\alpha$.

Let us assume that the correlation length, and hence the $\sigma$ susceptibility, 
$\chi_\sigma(m)$, are finite in the 
chiral limit. In such a case the flavor singlet scalar condensate behaves as 

\begin{equation}
\left< S\right>_{{\theta=0}_{\hskip 0.1cmm\rightarrow 0}} \approx \chi_{\sigma}\left(0\right) m
\label{condenmp}
\end{equation}
but since equation (\ref{mascuriosob}) tells us that 
$\left< S\right>_{Q=0} = \left< S\right>_{\theta=0}$, 
equation (\ref{condenmp}) holds also in the $Q=0$ sector,

\begin{equation}
\left< S\right>_{{Q=0}_{\hskip 0.1cmm\rightarrow 0}} \approx \chi_{\sigma}\left(0\right) m
\label{condenmpq0}
\end{equation}
The $Q=0$ sector is on the other hand free from the $U(1)_A$ global anomaly, hence the 
following relation between the flavor singlet scalar condensate 
$\left\langle S\right\rangle_{Q=0}$, and the $\eta$ and $\bar\pi$ susceptibilities, 
$\chi_\eta\left(m\right)_{Q=0}$, $\chi_{\bar\pi}\left(m\right)_{Q=0}$, 
holds in this sector\footnote{Notice that notwithstanding the $\sigma$ and $\bar\pi$ 
susceptibilities in the $Q=0$ sector are equal to the corresponding quantities in $QCD$ 
at $\theta=0$ in the thermodynamic limit, this is not true for the $\eta$ 
susceptibility, as discussed in reference \cite{trece}.}

\begin{equation}
\chi_{\bar\pi}\left(m\right)_{Q=0} = 
\chi_\eta\left(m\right)_{Q=0} = \frac{\left\langle S\right\rangle_{Q=0}}{m},
\label{chietaq0}
\end{equation}

Equations (\ref{condenmpq0}) and (\ref{chietaq0}) tell us that the flavor singlet 
scalar and pseudoscalar susceptibilities, in the $Q=0$ sector, take the same value, 
$\chi_{\sigma}\left(0\right)$, in the chiral limit, and this is a rather unexpected 
result because $\chi_\sigma\left(0\right)_{Q=0}-\chi_\eta\left(0\right)_{Q=0}$ is 
an order parameter for the spontaneously broken $U(1)_A$ symmetry.

A loophole to this paradoxical result would be a 
divergent flavor singlet scalar 
susceptibility, $\chi_\sigma\left(m\right)$, in the chiral limit. However it could 
also be that, for some accidental reason, the quark mass term selects an 
$\alpha$ -ground state, in the chiral limit, in which
$\chi_\sigma\left(0\right)_{Q=0}=\chi_\eta\left(0\right)_{Q=0}$. 
We will therefore continue exploring the physical consequences of assuming the 
correlation length, and $\chi_\sigma\left(m\right)$, are finite in the chiral limit.
 
The flavor singlet scalar 
$\left\langle S\left(x\right)S\left(0\right)\right\rangle_{Q=0}$ 
and pseudoscalar 
$\left\langle P\left(x\right)P\left(0\right)\right\rangle_{Q=0}$ 
correlation functions transform under $U(1)_A$ rotations of angle $\alpha$ as 

$$
\left\langle S\left(x\right)S\left(0\right)\right\rangle^{\alpha}_{Q=0} = 
\cos^2\alpha \left\langle S\left(x\right)S\left(0\right)\right\rangle^{\alpha=0}_{Q=0} +
\sin^2\alpha \left\langle P\left(x\right)P\left(0\right)\right\rangle^{\alpha=0}_{Q=0} +
$$
$$
\hskip 3cm\sin\alpha\cos\alpha\left(\left\langle S\left(x\right)P\left(0\right)\right
\rangle^{\alpha=0}_{Q=0} + \left\langle P\left(x\right)S\left(0\right)\right
\rangle^{\alpha=0}_{Q=0}\right)
$$

$$
\left\langle P\left(x\right)P\left(0\right)\right\rangle^{\alpha}_{Q=0} =
\sin^2\alpha \left\langle S\left(x\right)S\left(0\right)\right\rangle^{\alpha=0}_{Q=0} +
\cos^2\alpha \left\langle P\left(x\right)P\left(0\right)\right\rangle^{\alpha=0}_{Q=0} -
$$
\begin{equation}
\hskip 3cm\sin\alpha\cos\alpha\left(\left\langle S\left(x\right)P\left(0\right)\right
\rangle^{\alpha=0}_{Q=0} + \left\langle P\left(x\right)S\left(0\right)\right
\rangle^{\alpha=0}_{Q=0}\right)
\label{rotations}
\end{equation}
and therefore the flavor singlet
scalar and pseudoscalar susceptibilities, in the $Q=0$ sector, in the chiral limit, 
take the value 
$\chi_{\sigma}\left(0\right)$ not only in the ground state selected by the quark
mass term, but also in all the other $\alpha$-states, provided that parity is 
not spontaneously broken\footnote{Because  $SU(2)_A$ symmetry is fulfilled in the 
$Q=0$ sector in the chiral limit, the flavor singlet scalar and pseudoscalar 
condensates vanish in the $\alpha$-ground state selected by the quark-mass term, 
and therefore also in all other $\alpha$-states. Hence the disconnected contributions 
to the connected correlation functions are always canceled.}.

Moreover a simple calculation, based on an anomalous $U(1)_A$ 
transformation in the chiral limit, gives the following relation for the 
$\theta$-dependence of the flavor singlet scalar correlation function at vanishing 
quark mass 

\begin{equation}
\left\langle S\left(x\right)S\left(0\right)\right\rangle^{m=0}_{\theta} = \cos^2\left(
\frac{\theta}{2}\right)
\left\langle S\left(x\right)S\left(0\right)\right\rangle^{m=0}_{\theta=0} + \sin^2\left(
\frac{\theta}{2}\right)
\left\langle P\left(x\right)P\left(0\right)\right\rangle^{m=0}_{\theta=0}
\label{vayava}
\end{equation}
By performing the integral over the
$\theta$-angle in equation \ref{vayava}, as stated by equation \ref{mascurioso}, we get

\begin{equation}
\left\langle S\left(x\right)S\left(0\right)\right\rangle^{m=0}_{Q=0} = 
\frac{1}{2}
\left\langle S\left(x\right)S\left(0\right)\right\rangle^{m=0}_{\theta=0} + 
\frac{1}{2}
\left\langle P\left(x\right)P\left(0\right)\right\rangle^{m=0}_{\theta=0}
\end{equation}
and therefore a similar relation for the susceptibilities holds

\begin{equation}
\chi_{\sigma}\left(0\right)_{Q=0} = \frac{
\chi_{\sigma}\left(0\right) + \chi_{\eta}\left(0\right)}{2}
\label{vaya}
\end{equation} 

The flavor singlet scalar susceptibility of the $Q=0$ sector in the chiral 
limit is the average of this quantity over all $\alpha$-ground states 
\cite{pdc}, but we have previously shown that it takes the same value, 
$\chi_{\sigma}\left(0\right)$, in all $\alpha$-ground states. The compatibility of 
this result with equation (\ref{vaya}) requires therefore 
the $\eta$ and $\sigma$ susceptibilities to be equal, in contradiction with the 
assumption that the topological effects of the $U(1)_A$ anomaly survive in the high temperature 
phase. 

We conclude that the assumption on the finitude of the correlation length and 
$\sigma$-susceptibility 
in the chiral limit is wrong, this susceptibility diverges, and the free energy 
density is therefore singular at $m=0$. 

Under the standard assumption that the critical behavior of the model is 
well described by a power law for the flavor singlet scalar condensate 

\begin{equation}
\left< S\right>_{{\theta=0}_{\hskip 0.1cmm\rightarrow 0}} \approx C\left(T\right) 
m^{\frac{1}{\delta}}
\end{equation}
we get that the flavor singlet scalar susceptibility 
$\chi_\sigma$ diverges at any $T\ge T_c$ in the chiral limit as\footnote{The critical 
exponent $\delta$ should be that of the three-dimensional $O(4)$ vector 
universality class, $\delta = 4.789(6)$, at $T=T_c$. Universality arguments 
suggest it would be $T$-independent, but we do not exclude a temperature dependence 
of $\delta$ corresponding to a critical line with continuously varying critical 
exponents.}

\begin{equation}
\chi_\sigma\left(m\right) \approx C\left(T\right) \frac{1}{\delta} 
m^{\frac{1-\delta}{\delta}},
\end{equation}
and since the $SU(2)_A$ symmetry is not anomalous, the pion susceptibility verifies 
the relation 

\begin{equation}
\chi_{\bar\pi}\left(m\right) = \frac{\left\langle S\right\rangle}{m},
\label{chipi2}
\end{equation}
and diverges also in the chiral limit as
\begin{equation}
\chi_{\bar\pi}\left(m\right) \approx C\left(T\right) 
m^{\frac{1-\delta}{\delta}},
\end{equation}
Moreover the vector meson $\bar\delta$ susceptibility, $\chi_{\bar\delta}$, which 
is bounded by the scalar susceptibility, $\chi_\sigma$, verifies the 
following inequality

\begin{equation}
\chi_{\bar\pi}\left(m\right) - \chi_{\bar\delta}\left(m\right) \ge 
\chi_{\bar\pi}\left(m\right) - \chi_\sigma\left(m\right) \approx C\left(T\right)
\frac{\delta-1}{\delta}
m^{\frac{1-\delta}{\delta}}
\label{divergente}
\end{equation}
which shows explicitly that 
$\chi_{\bar\pi}\left(m\right) - \chi_{\bar\delta}\left(m\right)$ diverges in the chiral 
limit.

This result seems to be ruled out by the results of a numerical simulation of 
two-flavor QCD by Tomiya et al. reported in reference \cite{catorce}, 
where they find a value of 
$\chi_{\bar\pi}\left(m\right) - \chi_{\bar\delta}\left(m\right)$ 
in the chiral limit, and in a temperature range $T\sim 190-220$ MeV slightly above 
the critical 
temperature $T_c$, that not only does not diverge but is compatible with zero. However, previous
results by Dick et al. \cite{sharma} on larger lattices, but using overlap fermions only in the
valence sector, seem to predict a divergent
$\chi_{\bar\pi}\left(m\right) - \chi_{\bar\delta}\left(m\right)$ in the chiral
limit, in agreement with equation (\ref{divergente}).

\section{Phase diagram of QCD in the Q=0 topological sector}

The nonanalyticity of the free energy density at $m=0$ can also be shown by an 
alternative or complementary way. The $SU(2)_A$ symmetry is fulfilled in $QCD$ 
at any $T> T_c$, and therefore the up and down scalar condensates $\left< S_u\right>$, 
$\left< S_d\right>$ vanish in the chiral limit $m_u = m_d = 0$. However if we 
consider $QCD$ with two nondegenerate quark flavors, and take the limit 
$m_u\rightarrow 0$ keeping $m_d$ fixed, or vice versa, the condensate 
$\left< S_u\right>$, or $\left< S_d\right>$, takes a nonvanishing mean 
value due to the fact that the $U(1)_u$ symmetry at $m_u=0$, or the $U(1)_d$ symmetry 
at $m_d=0$, which would enforce the condensate to be zero, is anomalous. 
But since equation (\ref{mascuriosob}) can be applied to these condensates, 
this result tell us
that the $Q=0$ topological sector, which is not anomalous, spontaneously breaks the 
$U(1)_u$ axial symmetry at $m_u=0, m_d\ne 0$, and the $U(1)_d$ symmetry at 
$m_d=0, m_u\ne 0$. The phase diagram of $QCD$ in the $Q=0$ topological sector, in the 
$(m_u, m_d)$ plane, shows therefore two first order phase transition lines, which coincide 
with the coordinate axes, finishing at the end point $m_u=m_d=0$, which is a critical 
point for any $T> T_c$.

Equation (\ref{mascuriosob}) tell us that the critical equation of state of $QCD$ 
at $\theta=0$ should be the same as the one of the $Q=0$ topological sector, which 
should show a divergent correlation length at any $T>T_c$. We expect therefore a 
continuous finite temperature chiral transition, and a divergent correlation length
for any $T\ge T_c$, and 
because the symmetry breaking pattern is, in the two flavor model, 
$SU(2)_L\times SU(2)_R\rightarrow SU(2)_V$, the critical equation of state should 
be that of the three-dimensional $O(4)$ vector universality class \cite{piswil}, 
which shows a critical exponent $\delta = 4.789(6)$ \cite{hasen} ($\delta = 3$ 
in the mean field or Landau approach).
 
For $N_f\ge 3$ a similar argument on the phase diagram of the $Q=0$ sector applies, but the
scenario that emerges in this case is also not plausible because no stable fixed points 
are expected in the corresponding Landau-Ginzburg-Wilson $\Phi^4$ theory compatible 
with the given symmetry-breaking pattern \cite{ettore}.

\section{Conclusions and discussion}

The aim to elucidate the existence of new low-mass weakly interacting particles from 
a theoretical, phenomenological, and experimental point of view, is intimately related 
to the role of the $\theta$ parameter in $QCD$. Indeed the axion is one of 
the more interesting candidates to make the dark matter of the universe, and 
the $QCD$ axion model relates the topological susceptibility $\chi_T$ 
at $\theta=0$ with
the axion mass $m_a$ and decay constant $f_a$ through the relation 
$\chi_T = m^2_a f^2_a$, the 
axion mass being an essential ingredient in the calculation of the
axion abundance in the Universe. 
Moreover, the way in which the $U(1)_A$ anomaly manifests itself in the chiral
symmetry restored phase of $QCD$ at high temperature could be tested when probing 
the $QCD$ phase transition in relativistic heavy ion collisions.

With these motivations we started recently an investigation of the topological
properties of $QCD$ in the high temperature chiral symmetric phase in 
reference \cite{trece}. 
The starting hypothesis in \cite{trece} was to assume
that the perturbative expansion of the free energy density in powers of the quark
mass, $m$, has a nonvanishing convergence radius in the high temperature chiral
symmetric phase of $QCD$, which is just what we expect if all correlation lengths remain
finite in the chiral limit, and the spectrum of the model shows therefore a mass gap
also in this limit. The main conclusion in \cite{trece} was 
that all the topological effects of the axial anomaly should disappear in this phase,
the topological susceptibility and all $\theta$-derivatives of the free energy
density vanish, and the theory becomes $\theta$ independent at any
$T > T_c$ in the infinite-volume limit. Accordingly, the free energy density should be a
singular function of the quark mass, in the chiral limit, if the topological effects
of the $U(1)_A$ anomaly survive in the chiral symmetry restored phase of QCD at finite
temperature.

Ongoing with this research line, the main purpose of this article has been to further
investigate this issue. To this end our starting hypothesis has been now to assume
that the topological effects of the anomaly survive in the high temperature phase of
$QCD$, and the model shows therefore a
nontrivial $\theta$-dependence in this phase. Under this assumption we have shown
that indeed, the free energy density is a singular function of the quark mass, $m$,
in the chiral limit at any $T>T_c$, and that the correlation length and the
$\sigma$ and $\bar\pi$ susceptibilities diverge in this limit.
Under the same assumption we have also shown that the difference between the
$\bar\pi$ and $\bar\delta$ susceptibilities diverges in the chiral limit at any
$T\ge T_c$, a result which seems to be excluded by recent results of 
Tomiya et al. \cite{catorce} from numerical simulations of two-flavor QCD, thus 
suggesting the topological effects of the $U(1)_A$ anomaly are absent in the chiral
symmetric phase of two-flavor $QCD$. However, previous
results by Dick et al. \cite{sharma} on larger lattices, but using overlap fermions only in the
valence sector, seem to predict a divergent
$\chi_{\bar\pi}\left(m\right) - \chi_{\bar\delta}\left(m\right)$ in the chiral
limit, in agreement with equation (\ref{divergente}), and any further clarification on this
point would be therefore welcome.

We have also discussed that the previous results for the two-flavor model apply also to 
$N_f\ge 3$. However, universality and 
renormalization-group arguments, based on the most general Landau-Ginzburg-Wilson 
$\Phi^4$ theory compatible with the given symmetry-breaking pattern, make this scenario 
not plausible too because no stable fixed points 
are expected in the corresponding Landau-Ginzburg-Wilson $\Phi^4$ theory 
for $N_f\ge 3$ \cite{ettore}.

\section{Acknowledgments}

We acknowledge Sayantan Sharma for clarifying us some details of his work \cite{sharma}. 
This work was funded by Ministerio de
Economía y Competitividad under Grant No. FPA2015-65745-P (MINECO/FEDER).

\vfill
\eject


\begin{thebibliography}{99}
  
\bibitem{vicari}
E. Vicari, H. Panagopoulos, 
\textit{Phys. Rep.} \textbf{470}, 93 (2009).

\bibitem{weinberg}
S. Weinberg,F. Wilczek, 
\textit{Phys. Rev. Lett.} \textbf{40}, 223 (1978).

\bibitem{wilczek}
F. Wilczek, 
\textit{Phys. Rev. Lett.} \textbf{40}, 279 (1978).

\bibitem{pq}
R.D. Peccei, H.R. Quinn, 
\textit{Phys. Rev. Lett.} \textbf{38}, 1440 (1977); {Phys. Rev.} \textbf{D16}, 
1791 (1977).

\bibitem{martinelli}
C. Bonati, M. D’Elia, M. Mariti, G. Martinelli, M. Mesiti, F. Negro, F. Sanfilippo,
G. Villadoro, 
\textit{J. High Energy Phys.} \textbf{03}, (2016) 155.

\bibitem{petre}
P. Petreczky, H-P. Schadler and S. Sharma, 
\textit{Phys. Lett. B} \textbf{762}, 498 (2016).

\bibitem{javier}
Sz. Borsanyi et al., \textit{Nature (London)} \textbf{539 }, 69 (2016).

\bibitem{cohen1}
T. D. Cohen, 
\textit{Phys. Rev. D} \textbf{54}, R1867 (1996).

\bibitem{cohen2}
T. D. Cohen, 
\textit{ arXiv:nucl-th/9801061}.

\bibitem{cuatro}
C. Bernard, T. Blum, C. DeTar, S. Gottlieb, U. M. Heller,
J. E. Hetrick, K. Rummukainen, R. Sugar, D. Toussaint, and
M. Wingate, 
\textit{Phys. Rev. Lett.} \textbf{78}, 598 (1997).

\bibitem{cinco}
S. Chandrasekharan, D. Chen, N. H. Christ, W. -J. Lee, R. Mawhinney and P. M. Vranas,
\textit{Phys. Rev. Lett.} \textbf{82}, 2463 (1999).

\bibitem{seis}
H. Ohno, U. M. Heller, F. Karsch and S. Mukherjee, 
\textit{Proc.Sci., LATTICE2011} (2011) 210.

\bibitem{siete}
A. Bazavov et al. [HotQCD Collaboration], 
\textit{Phys. Rev. D} \textbf{86}, 094503 (2012).

\bibitem{ocho}
T. G. Kovacs and F. Pittler, 
\textit{Proc.Sci., LATTICE2011} (2011) 213.

\bibitem{nueve}
G. Cossu, S. Aoki, S. Hashimoto, T. Kaneko, H. Matsufuru, J. -i. Noaki and E. Shintani, 
\textit{Proc.Sci., LATTICE2011} (2011) 188.

\bibitem{diez}
S. Aoki, H. Fukaya and Y. Taniguchi, 
\textit{Phys. Rev. D} \textbf{86}, 114512 (2012).

\bibitem{once}
G. Cossu, S. Aoki, H. Fukaya, S. Hashimoto, T. Kaneko and H. Matsufuru, 
\textit{Phys. Rev. D} \textbf{87}, 114514 (2013); \textbf{88},
019901 (2013).

\bibitem{sharma}
V. Dick, F. Karsch, E. Laermann, S. Mukherjee, and S. Sharma, 
\textit{Phys. Rev. D} \textbf{91}, 094504 (2015).

\bibitem{doce}
G. Cossu et al.,
\textit{Proc.Sci., LATTICE2015} (2016) 196.

\bibitem{yamamoto}
T. Kanazawa and N. Yamamoto, 
\textit{J. High Energy Phys.} \textbf{01}, (2016) 141.

\bibitem{docebis}
B.B. Brandt, A. Francis, H.B. Meyer, O. Philipsen, D. Robainad and H. Wittig,
\textit{J. High Energy Phys.} \textbf{12}, (2016) 158.

\bibitem{trece}
V. Azcoiti, 
\textit{Phys. Rev. D} \textbf{94}, 094505 (2016).

\bibitem{catorce}
A. Tomiya, G. Cossu, S. Aoki, H. Fukaya, S. Hashimoto, T. Kaneko and J. Noaki, 
\textit{arXiv:1612.01908 [hep-lat]}.

\bibitem{ding}
H.-T. Ding, 
\textit{arXiv:1702.00151 [hep-lat]}.

\bibitem{Ginsparg}
P.H. Ginsparg and K. G. Wilson,
\textit{Phys. Rev. D} \textbf{25}, 2649 (1982).

\bibitem{Neuberger}
H. Neuberger,
\textit{Phys. Lett. B}\textbf{417} 141 (1998); 
\textbf{427} 353 (1998).

\bibitem{Luscher}
M. Luscher, \textit{Phys. Lett. B} \textbf{428} 342 (1998).

\bibitem{Victor}
P. Hasenfratz, V. Laliena and F. Niedermayer,
\textit{Phys. Lett. B} \textbf{427}, 125 (1998).

\bibitem{pdc}
V. Azcoiti, V. Laliena and X.Q. Luo, 
\textit{Phys. Lett. B} \textbf{354}, 111 (1995).

\bibitem{piswil}
R.D. Pisarski and F. Wilczek,
\textit{Phys. Rev. D} \textbf{29}, 338 (1984).

\bibitem{hasen}
M. Hasenbusch, 
\textit{J. Phys. A} \textbf{34} 8221 (2001).

\bibitem{ettore}
E. Vicari, 
\textit{Proc.Sci., LATTICE2007} (2007) 023.

\end{thebibliography}
\end{document}